\def\alt{\ {\raise -0.25em \hbox{{$\buildrel < \over \sim $ }}}\ }
\def\agt{\ {\raise -0.25em \hbox{{$\buildrel > \over \sim $ }}}\ }

\let\t=\theta
\documentstyle[prb,twocolumn,aps,epsf]{revtex}
\begin{document}
\draft
\input amssym.def
\input amssym
\twocolumn[\hsize\textwidth\columnwidth\hsize\csname @twocolumnfalse\endcsname
\title{Leading Temperature Corrections to Fermi Liquid Theory in 
Two Dimensions}
\author{Gennady Y. Chitov and Andrew J. Millis}
\address{Center for Materials Theory}
\address{Department of Physics and Astronomy}
\address{Rutgers University, Piscataway, New Jersey 08854}
\maketitle

\begin{abstract}
We  calculate the basic parameters of the Fermi Liquid: the scattering 
vertex, the Landau interaction function, the effective mass, 
and physical susceptibilities 
for a model of two-dimensional (2D) fermions with a short ranged 
interaction at non-zero temperature. 
The leading temperature dependences of the spin components of the 
scattering vertex, the Landau function, and the spin
susceptibility are found to be linear. 
$T$-linear terms in the effective mass and in the ``charge-sector''-
quantities are found to cancel to second order in the interaction, 
but the cancellation is argued not to be generic. 
The connection with previous studies 
of the 2D Fermi-Liquid parameters is discussed.
\end{abstract} 


]
%
\narrowtext
%
%
The question of the low-energy behavior of 
two-dimensional (2D) Fermi Liquid (FL) is of long-standing and of 
fundamental importance. One important motivation has been the non-Fermi-Liquid 
behavior observed in high-$T_c$ superconductors above $T_c$.\cite{HighTc} 
Interest has grown 
in recent years also because these leading corrections provide
the ``bare'' temperature dependence of the parameters in theories describing
quantum critical phenomena in metals.\cite{HeM} Indeed
a number of surprising experimental results\cite{CeCu6,Lonzarich} have been
argued\cite{Coleman98,Si00} to imply an unusal underlying temperature,
momentum or frequency dependence of electronic susceptibilities.

Rather surprisingly, the issue of the leading temperature corrections to Fermi 
Liquid Theory (FLT)  remains controversial at certain points. 
For example, the textbook  Sommerfeld expansion suggests
that physical quantities may in general be expanded in powers of $T^2$.
However, this is known not to be true.
In particular, it was found\cite{Amit67} 
that the leading temperature correction to the specific heat
coefficient $\gamma=C/T$ was $T^{2}\ln T$ in d=3 spatial 
dimensions and $T$ in d=2.\cite{Bedell93}.
Whether the spin and charge susceptibilities display
similarly anomalous (i.e., non-$T^2$) temperature dependence 
is a subject of a contradictory literature:
see, e.g., Ref.~[\onlinecite{CP77}], discussion and references there in. 
For a most recent reassesment of such results see Ref.~[\onlinecite{BK97}].
The prevailing conclusion  was that of 
Carneiro and Pethick\cite{CP77} who found no leading 
$T^2 \ln T$ correction to the spin susceptibility of the 3D FL.
Their arguments imply that terms $\propto T$ are absent in 2D.

The heuristic argument  for the absence of anomalous terms in $T$ 
or in $q$  in response functions is that although these terms are 
known to occur in individual diagrams, they cancel in physical quantities 
due to Ward identities. We note that in the existing literature on 
this point it is assumed that the crucial coupling is  between 
quasiparticles and long-wavelength collective modes.
However the possibility of ``$2k_{F}$ singularities'',
i.e., anomalous temperature terms coming from processes involving large 
($\sim 2k_{F}$) momentum transfers,  have been discussed in the context 
of semiconductor physics.  Stern was the first to note\cite{Stern80} 
that in a 2D electron gas the electron scattering rate was 
proportional to $T$ due to $2k_{F}$ effects. The consequences of 
the $2k_{F}$ effects for the leading $T$-dependence of 2D FL quantities
seem not to have been considered in the literature.
In this paper we present an analysis taking into account both 
$2k_{F}$ effects and Ward identities.

The issue of the leading correction to FLT have recently been revived by 
two different papers. 
Belitz, Kirkpatrick and Vojta\cite{BK97} presented perturbative
calculations, mode-coupling arguments and power counting estimates
which showed that the leading $q$ dependence of the spin susceptibility (but
not the charge susceptibility) was $|q|$ in 2D  and $q^{2}\ln q$ in 3D.
They did not find the analogous $T$-correction explicitly, but 
concluded however that one should generally expect a linear $T$-term 
in the 2D FL susceptibility ($T^2\ln T$ in 3D). This dependence has 
important implications  for the theory of the quantum
critical metallic ferromagnet.\cite{BeMi} They also focused on 
long-wavelength contributions.

In the other study, S{\'e}n{\'e}chal and one of us\cite{GD98} predicted the occurence 
of the linear $T$-corrections to the FL vertices from one-loop Renormalization 
Group (RG) calculations based on a 2D effective action.
However, the behavior of other FL quantities was not determined.

To elucidate the question  in the most transparent way, we apply  
perturbation theory for 2D contact-interacting spin-$\frac12$ fermions, 
starting from a microscopic action. Although Landau FLT is not a perturbative 
theory, for sufficiently weak repulsive interactions  
one should be able to find  the parameters of a  stable FL in terms of  
coupling series. Since both papers\cite{BK97,GD98} predict the linear
$T$-terms appearing in the second order of the {\it effective}
interaction, this effect should be seen perturbatively
at second order in the miscroscopic interaction coupling.
We present what is apparently the first exact calculation of the leading 
$T$-dependence of the effective mass, Landau parameters and response functions
of a 2D electron gas, to second order in the interaction
strength, including all channels and all momentum processes.
%
%
%
%

$\bullet$ {\bf The model:}
We treat interacting fermions at finite temperature in the
standard path integral Grassmannian formalism.\cite{Negele88}
The partition function is given by the path integral
$
{\cal Z} =\int {\cal D}\bar\psi{\cal D}\psi~{\rm exp}(S_0 + S_{\rm int})
$,
where
\begin{equation}
\label{S0}
S_0 = \int_{({\bf 1})}\bar\psi_{\alpha}({\bf 1})\left[
i\omega_1+\mu-\epsilon({\bf k}_1)\right]\psi_{\alpha}({\bf 1})
\end{equation}
We have adopted the condensed notations:
$({\bf i}) \equiv ({\bf k}_i,\omega_i)$ and 
$\int_{({\bf i})}\equiv{1\over\beta}\int{d{\bf k}_i\over(2\pi)^2}
\sum_{\omega_i}$,
$\beta$ is the inverse temperature, $\mu$ the chemical potential,
$\omega_i$ the fermion Matsubara frequencies and $\psi_{\alpha}({\bf i})$ a
two-component Grassmann field with a spin  index $\alpha$.
Summation over repeated indices is implicit throughout this paper.
We set $k_B=1$ and $\hbar=1$.
We consider mainly electrons with the bare spectrum of a free gas 
$\epsilon({\bf k})=k^2/2m$ and  the circular Fermi surface, 
but discuss the consequences of generic spectra and a
non-circular Fermi surface below. We take
\begin{eqnarray}
\label{SintHub}
S_{\rm int} =- {u \over 4 \nu_0} \int_{({\bf 1},..,{\bf 4})}
&~& \bar\psi_{\alpha}({\bf 1})\bar\psi_{\beta}({\bf 2})
\psi_{\gamma}({\bf 3})\psi_{\varepsilon}({\bf 4}) 
T^{\alpha\beta}_{\gamma\varepsilon} \nonumber \\
&\times& \delta^{(2+1)}({\bf 1}+{\bf 2}-{\bf 3}-{\bf 4})
\end{eqnarray}
Here $u>0$ corresponds to repulsion, $\nu_0=m/2\pi$
is the free 2D density of states per spin, 
the spin antisymmetric operator
$T^{\alpha\beta}_{\gamma\varepsilon}~\equiv~ 
\delta_{\alpha\varepsilon}\delta_{\beta\gamma} -
\delta_{\alpha\gamma}\delta_{\beta\varepsilon}$.
For this model we calculate the vertices, effective mass,
and the uniform response functions.
%
%
%
%

$\bullet$ {\bf Four-point 1PI vertex and the FL vertices:}
The 1PI vertex  $\hat \Gamma({\bf 1},{\bf 2};{\cal Q})$
is defined 
in the standard way (see, e.g., Ref.[\onlinecite{Negele88}]),
where the  transfer vector
$
{\cal Q} = ({\bf q},\Omega) 
$,
and $\Omega $ is a bosonic Matsubara frequency.  
To shorten notations we denote operators in spin space with a 
circumflex.
To find the FL vertices  we need to calculate 
$\hat \Gamma({\bf 1},{\bf 2};{\cal Q})$ in the limit of zero transfer
${\cal Q}$ with incoming momenta lying  on the Fermi surface. 
Since the Fermi surface is circular,  
the vertex can be parametrized by the relative angle between incoming
momenta. It is known\cite{AGD,Lifshitz80}
that the limit ${\cal Q} \to 0$ is not 
unique.
We define two  vertices $\hat \Gamma^q(\t_{12}) \equiv 
\hat \Gamma(\theta_{12};{\bf q} \to 0,\Omega =0)$ 
and $\hat \Gamma^\Omega(\t_{12}) \equiv 
\hat \Gamma(\theta_{12};{\bf q}=0,\Omega \to 0)$,  
which can be related to the components of the physical 
scattering vertex ($A,~B$) 
and the Landau interaction function ($F,~G$), respectively. 
Namely:\cite{Lifshitz80}
\begin{equation}
\label{scat-comp}
-2 \nu_R Z^2  \Gamma^{\alpha \beta (q)}_{\gamma \varepsilon}
=A \delta_{\alpha \gamma }\delta_{\beta \varepsilon }
+B \sigma_{\alpha \gamma }^a \sigma_{\beta \varepsilon }^a
\end{equation}
where $\nu_R=m^{\ast}/2\pi$,
$Z$ is the field renormalization constant,
$\hat \sigma^a$ are  Pauli's matrices.
Two components ($F,~G)$ of the Landau function are defined by 
analogous equation, with the substitution
$q \mapsto \Omega$, $A \mapsto F$, $B \mapsto G$.
%
%
%
%

$\bullet$ {\bf Scattering vertex and Landau function:}
The one-loop approximation 
for ${\hat\Gamma}({\bf 1},{\bf 2};{\cal Q})$ 
in diagrammatic form is given in Fig.~1. 
In this approximation we calculate 
the FL vertices (scattering vertex, Landau function) using definition
(\ref{scat-comp}).
At the one-loop level 
we can put $\nu_R=\nu_0$ ($m^{\ast}=m$) and $Z=1$.
By doing the direct analytic evaluation of the each diagram's 
contribution to the vertex we find the Fourier components of the angular-dependent
FL vertices  in terms of the temperature series. This series comes
from the contributions of the ZS' and BCS loops.
The details will be given in a companion paper.\cite{GAP1} 
We find the leading temperature corrections 
to first two Fourier components of the vertices:
\begin{eqnarray}
\label{chspdT}
\delta A_0=\delta F_0=\delta A_1=\delta F_1=
-u^2{\pi^2 \over 24}{T^2 \over E_F^2} \\
\delta B_0=\delta G_0=-\delta B_1=-\delta G_1=
-u^2{T \over E_F}\ln2
\end{eqnarray}
We should mention that taken separately, each contribution of the
ZS'- or BCS bubble gives a leading $T$-term
to the Fourier components of the vertices.  
However, a cancellation of such terms coming from 
two graphs occurs in the ``charge sector'' (i.e., in $A$, $F$ components), 
while the linear $T$-terms survive in the ``spin sector'' 
($B$, $G$ components). The temperature dependence of the ZS' 
contribution to the Fourier components of the vertices comes 
from integration  around the ``effective transfer'' through the loop  
$|{\bf k}_1-{\bf k}_2| \sim 2k_F$, i.e., when incoming momenta 
${\bf k}_1 \sim -{\bf k}_2$.  
In the same vein, the temperature dependence of the BCS 
contribution into 
the Fourier components of the vertices comes from regions of small 
${\bf k}_1+{\bf k}_2$, 
i.e., again when ${\bf k}_1 \sim -{\bf k}_2$. In other words, the 
temperature dependence comes from what we called previously
``$2k_F$-effects''.

We expect that the cancellation of the $T$-linear terms in the 
charge sector of the vertices (\ref{chspdT}) is an artefact  
of our simple model calculation in which all three one-loop terms 
have the same factor $u^2$ in front of the bubble contributions. 
Had we had a bare coupling 
{\it function} of, say, two incoming momenta and transfer, then 
the coupling factors would have been different
in each of the three graphs, and, the anomalous  $T$-corrections 
would not have cancelled.

The linearity of the leading $T$-corrections to the
vertices  seems to be generic. The same  $T$-dependence (apart from presumably 
model-sensitive prefactors) was obtained in the previous RG analysis\cite{GD98} 
of the {\it effective action} for 2D spinless fermions with a 
linearized one-particle spectrum and a momentum-dependent coupling function. 
According to Ref.~[\onlinecite{BK97}] such temperature behavior 
can be understood from dimensional arguments.

%
%
%
%

$\bullet$ {\bf Effective mass:} It is defined by the following equation
\begin{equation}
\label{mstara}
{m^{\ast} \over m}={ 1- 
{\partial \Sigma({\bf 1}) \over \partial i \Omega} 
\over 
1+{m \over k_F} {\partial \Sigma({\bf 1}) \over \partial {\bf k} }
{{\bf k}_1  \over k_F}  } 
\Bigg\vert^{ \omega_1 = 0}_{{\bf k}_1 \in S_F}
\end{equation}
To second order we have 
\begin{equation}
\label{mstarappr}
{m^{\ast} \over m}=1-
\Bigg[ 
{\partial \Sigma({\bf 1}) \over \partial  \omega}
+
{m \over k_F} {\partial \Sigma({\bf 1}) \over \partial {\bf k} }
{{\bf k}_1  \over k_F}  
\Bigg]
\Bigg\vert^{ \omega_1 = 0}_{{\bf k}_1 \in S_F}
+{\cal O}(u^3)
\end{equation}
Using then two Ward identities following from the charge conservation
and Galilean invariance,\cite{AGD,Lifshitz80} the above equation can be 
written as:
\begin{equation}
\label{mstarWIap}
{m^{\ast} \over m}=1
- \frac12 \int_{({\bf 2})}{ {\bf k}_1 {\bf k}_2 \over k_F^2}
\Gamma^{\alpha \beta}_{\alpha \beta}({\bf 1},{\bf 2}; \Omega \to 0) 
 \Delta({\bf 2})
\Big\vert^{ \omega_1 = 0}_{{\bf k}_1 \in S_F}
\end{equation}
where 
\begin{equation}
\label{Delta1}
\Delta({\bf n})={\beta \over 4}
{ \delta(\omega_n-\xi_{{\bf k}_n}) \over 
\cosh^2( \frac12 \beta \xi_{{\bf k}_n}) }
\end{equation}
Within our accuracy we can use the one-loop approximation for the 
vertex in Eq.~(\ref{mstarWIap}). One can easily verify that in the the 
zero-temperature limit Eq.~(\ref{mstarWIap})
recovers the standard result of the FLT,\cite{Lifshitz80} i.e., 
$
m^{\ast}(T=0)/  m=1+F_1(T=0) 
$.
A straightforward extension of this relationship  to finite temperatures 
like $m^{\ast}(T)/ m=1+F_1(T)$ is 
not valid since according to Eq.~(\ref{mstarWIap}) $m^{\ast}(T)$ contains 
an extra contribution from the ``off-shell'' integration over 
$k_2~(\xi_{{\bf k}_2})$ normal to the Fermi surface, albeit the factor 
$\beta / \cosh^2(\beta \xi_{{\bf k}_2}/2 )$ makes this contribution 
well-localized near the Fermi surface. In other words the vertex entering 
the r.h.s. of Eq.~(\ref{mstarWIap}) is not exactly the FL vertex $F(T)$  
(up to the normalization factor) as it is defined in the FLT, since one of its  
momenta (namely, ${\bf k}_2$) 
is not confined to the Fermi surface.   
After calculations we find that the linear-temperature terms, 
coming essentially from two one-loop contributions (ZS', BCS) to the 
vertex, cancel, resulting in  
\begin{equation}
\label{msFin}
{m^{\ast} \over m}=1+ \frac12 u^2 +{\cal O}(u^2T^2)
\end{equation}
In a close analogy with the cancellation of the linear-temperature terms
in the Fourier components of the FL vertices $A$ ($F$), here the 
cancellation occurs between additive linear-$T$ corrections coming 
from both ``on-shell'' (i.e., linear $T$-term coming from the 
$2k_F$-contribution to the vertex) and ``off-shell'' (i.e., 
the small-momentum contribution) integrations in two diagrams.
Moreover, the ``on-shell'' (``off-shell'') $T$-term of the ZS' graph 
cancells  the ``on-shell'' (``off-shell'') $T$-term of the BCS graph, 
correspondingly. We expect the cancellation does not occur at higher 
orders in the interaction. We have also evaluated (\ref{mstarappr}) 
for a generic 2D Fermi surface without explicitly using Ward 
identities, finding a $T$-linear term.\cite{GAP1} The result 
may be expressed as the sum of two terms, one arising from $2k_F$
processes and one from long-wavelength processes. The two 
contributions cancel for a circular Firmi surface, but not 
generically.

Calculations of the order $u^2$ term in the free energy show that 
analogous cancellations occur and there is no $T$-linear term 
in the specific heat coefficient $\gamma$, contrary to the results 
of Ref.[\onlinecite{Bedell93}].
%
%
%
%

$\bullet$ {\bf Response functions:} Using the same Ward 
identities as in the effective mass calculation, we found for the dynamic 
zero-transfer limit ($\Omega=0$, ${\bf q} \to 0$) of the density response 
function:
\begin{eqnarray}
\label{kapq4}
\varkappa &=& {m \over  \pi} \Big\{   1+u^2 +f_1-f_0 \Big\} \\
f_1 &\equiv&- { \pi \over m}  
\int_{({\bf 1},{\bf 2})} {{\bf k}_1 {\bf k}_2 \over k^2_1}
\Delta({\bf 1}) 
\Gamma^{\alpha \beta}_{\alpha \beta}({\bf 1},{\bf 2}; \Omega \to 0) 
\Delta({\bf 2}) \nonumber \\
f_0 &\equiv& - { \pi \over m}  \int_{({\bf 1},{\bf 2})}
\Delta({\bf 1})
\Gamma^{\alpha \beta}_{\alpha \beta}({\bf 1},{\bf 2}; \Omega \to 0) 
 \Delta({\bf 2}) \nonumber   
\end{eqnarray}
At $T=0$ we can read off from Eq.~(\ref{kapq4})    
$
\varkappa(T=0)= {m \over  \pi}(1+F_0^2+F_1-F_0)
$
which is nothing but the FLT result\cite{Lifshitz80} 
$
\varkappa^{FLT}= {m \over  \pi}{ 1+F_1 \over 1+F_0}
$,
expanded  up to the third order over the interaction. Adding into consideration 
the Ward identity following from the total spin conservation, we derived for
the uniform spin susceptibility (for details see Ref.~[\onlinecite{GAP1}]):
\begin{eqnarray}
\label{chiq4}
{\cal X}&=& {m {\goth g}^2 \over 4 \pi} \Big\{ 1+u^2 +f_1- g_0 \Big\} \\
 g_0 &\equiv& -{ \pi \over 3 m}  \int_{({\bf 1},{\bf 2})}
\Delta({\bf 1}) 
\sigma^a_{\gamma \varepsilon} \sigma^a_{\beta \alpha} 
\Gamma^{\alpha \varepsilon}_{\beta \gamma}({\bf 1},{\bf 2}; \Omega \to 0)  
\Delta({\bf 2}) \nonumber 
\end{eqnarray}
where ${\goth g}$ stands for the gyromagnetic ratio.
Once again, one can see that in the zero-temperature limit the above equation 
gives 
$
{\cal X}(T=0)={m {\goth g}^2 \over 4 \pi}(1+G_0^2+F_1-G_0)
$
reproducing thus the second-order expansion of the 
the FLT result\cite{Lifshitz80} 
$
{\cal X}^{FLT}={m {\goth g}^2 \over 4 \pi}{ 1+F_1 \over 1+G_0}
$.

We were able to analytically calculate the integrals on the r.h.s. of 
Eqs.~(\ref{kapq4},\ref{chiq4}) in the leading order of their
temperature dependence. 
We found that the leading linear $T$-corrections, which can be traced back 
to the ZS'- and BCS-loop contributions to the vertex, cancel in each
of the integral terms $f_0$ and $f_1$ in Eq.~(\ref{kapq4}) separately. The 
result for the density response is:
\begin{equation}
\label{kapfin}
\varkappa= 
{m \over \pi}(1-u-\frac12 u^2+u^2\ln{2\Lambda \over k_F})+{\cal O}(u^2T^2)
\end{equation}
where $\Lambda \gg k_F$ is the ultraviolet cutoff
we introduced to regularize the BCS loop.
Let us finally remind\cite{Lifshitz80} that the compressibility
$K=\varkappa/ n^2$, where $n$ is electron density. 

We may calculate the spin susceptibility in the same way. In this case
the second integral term $g_0$ in Eq.~(\ref{chiq4}) does not contain 
the contribution of the ZS' loop, so the linear $T$-term coming from the 
BCS loop survives. Thus the spin
susceptibility has  a linear leading temperature correction:
\begin{equation}
\label{sus}
{ {\cal X}(T) \over  {\cal X}(0) } \approx 1+u^2 {T \over E_F}
\end{equation}
where
$
{\cal X}(0) = {m {\goth g}^2 \over 4 \pi}
[ 1+u-u^2 ( \ln{2 \Lambda \over k_F} -\frac32 ) ]
$,
and the first omitted term is ${\cal O}(u^2T^2)$.

It is useful to keep in mind that albeit the response functions in 
Eqs.~(\ref{kapq4},\ref{chiq4}) are explicitly expressed in terms of 
the vertex only, those contributions indeed entangle both ``purely vertex'' 
corrections and self-energy corrections. The latter 
are just expressed in terms of the vertex via the Ward identities.

Let us come back to the argument for the cancellation of temperature 
terms in the response functions, appealing to the Ward identities.
These identities should work at each level in terms of the 
coupling expansion.  We have calculated the vertices at the one-loop level 
${\cal O}(u^2)$. Through the Ward identities  the self-energy correction 
were taken into account with the same accuracy. 
There are no more terms of the order ${\cal O}(u^2)$ to cancel the 
temperature dependence (\ref{sus}). Thus, the linear temperature dependence 
of susceptibility (or weaker temperature dependence of the compressibility) 
does not contradict the {\it exact} 
Ward identities known to us, moreover in our results 
for the response functions both vertex and self-energy corrections 
are included on the same footing {\it by using the Ward identities}.

%
%
%
%

$\bullet$ {\bf Conclusions:}
In this paper we have systematically examined the leading temperature 
corrections to FLT in two spatial dimensions.
Our results reveal the crucial importance of $2k_F$ processes mainly
neglected by other workers. We find for a 2D Galilean-invariant FL
that to order $u^2$ the leading $T$-dependence of the parameters in 
the spin sector is $T$, for the others it is $T^2$. 

We find that the standard relationship between the effective 
mass and the Landau parameter in a Galilean-invariant FL 
is violated by finite-temperature terms. The structure of the formulas
shows that those terms arise from states very near to the Fermi level 
(``off-shell'' quantities) beyond the scope of the classical FLT 
derivations.  

The particularly interesting new result we found is the leading linear 
temperature dependence of the spin susceptibility (\ref{sus}). 
According to the perturbative calculations of Belitz {\it et al},\cite{BK97}
the 2D FL susceptibility has a leading linear correction in 
$|{\bf q}|$ at $T=0$
with a positive coefficient which is of the second 
order in interaction, i.e., their result has a structure of 
Eq.~(\ref{sus}). Comparison between our Eq.~(\ref{sus})
and the prediction of Ref.[\onlinecite{BK97}] for ${\cal X}({\bf q},T=0)$
meets  heuristic expectations of a reciprocity between 
small $T$- and ${\bf q}$-dependencies. 

For more realistic models of electrons in (quasi)-2D crystals, i.e., 
for various tight-binding spectra and fillings, 
the free-gas-like square-root $2k_F$ 
singularities (with $k_F$ depending on a chosen direction in ${\bf q}$-space) 
are known to exist in the Lindhard functions.\cite{Sher93}
We think this is enough to result in linear $T$-terms in physical 
quantities analogous to what we found in this study.
We argue  that the rather accidential cancellation of the $T$-terms 
in some FL parameters is special to second order perturbation theory
and a circular Fermi surface, while the leading linear temperature 
corrections are a generic feature of the 2D FL.

We hope our results may be experimentally tested in real 2D FL systems.
For example, a very naive fit of the temperature dependence of the spin 
susceptibility in ${\rm Sr_2RuO_4}$ system\cite{Maeno} when it is in the 
2D metallic regime (above 3D crossover temperature) shows that the data are  
compatible with the form (\ref{sus}). 
We hope our results stimulate a more detailed
examination of the leading temperature dependences
of response functions in 2D systems.
 
%
%

$\bullet$ {\bf Acknowledgements:}
Stimulating conversations with S. Lukyanov are gratefully acknowledged. 
We thank V. Oudovenko for his help with 
the numerical tests of our results. This work is supported
by NSF DMR0081075.
%

%

\begin{figure}
\vglue 0.4cm\epsfxsize 08cm\centerline{\epsfbox{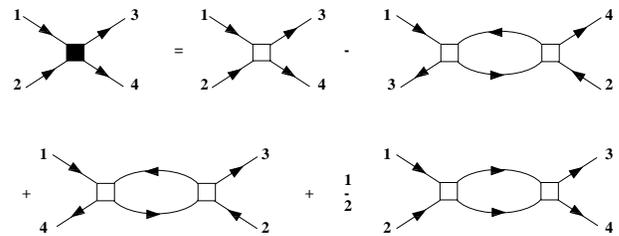}}\vglue 0.4cm
\caption{
Diagrammatic equation for  the four-point vertex at one-loop level.
The one-loop graphs are called ZS, ZS', and BCS in the order they appear 
on the r.h.s. of this equation.}
\end{figure}


\begin{references}
%
\bibitem{HighTc} For a recent comprehensive review and references see, e.g., 
T. Timusk and B. Statt, Rep. Prog. Phys. {\bf 62}, 61 (1999). 
%
\bibitem{HeM} J.A. Hertz, Phys. Rev. B {\bf 14}, 1165 (1976);
A.J. Millis,  Phys. Rev. B {\bf 48}, 7183 (1993).
%
\bibitem{CeCu6} H. von L{\"o}hneysen, {\it et al}, 
Phys. Rev. Lett. {\bf 72}, 3262 (1994).
%
\bibitem{Lonzarich} S.R. Julian, {\it et al}, 
JMMM {\bf 177-181}, 265 (1998).  
%
\bibitem{Coleman98} A. Schr{\"o}der, G. Aeppli, E. Bucher, R. Ramazashvili,
and P. Coleman,  Phys. Rev. Lett. {\bf 80}, 5623 (1998).
%
\bibitem{Si00} Q. Si, S. Rabello, K. Ingersent, and 
J.L. Smith, cond-mat/0011477. 
%
\bibitem{Amit67} D.J. Amit, J.W. Kane, and H. Wagner,
Phys. Rev. Lett. {\bf 19}, 425 (1967).
%
\bibitem{Bedell93} D. Coffey and K.S. Bedell, 
Phys. Rev. Lett. {\bf 71}, 1043 (1993).
%
\bibitem{CP77} G.M. Carneiro and C.J. Pethick,
Phys. Rev. B {\bf 16}, 1933 (1977).
%
\bibitem{BK97} D. Belitz, T.R. Kirkpatrick, and T. Vojta, 
Phys. Rev. B {\bf 55}, 9452 (1997).
%
\bibitem{Stern80} F. Stern, Phys. Rev. Lett. {\bf 44}, 1469 (1980).
%
\bibitem{BeMi} D. Belitz, T.R. Kirkpatrick, A.J. Millis,
and T. Vojta, Phys. Rev. B {\bf 58}, 14155 (1998).
%
\bibitem{GD98} G.Y. Chitov and D. S{\'e}n{\'e}chal, Phys. Rev. B  {\bf 57},
1444 (1998).
%
\bibitem{Negele88} J.W. Negele and H. Orland, 
{\it Quantum Many-Particle Systems} (Addison-Wesley, New York, 1988).
%
\bibitem{AGD} A.A. Abrikosov, L.P. Gorkov, and I.E. Dzyaloshinski, 1963, 
{\it Methods of Quantum Field Theory in Statistical Physics} (Dover, New York).
%
\bibitem{Lifshitz80} E.M. Lifshitz, L.P. Pitayevskii, 1980, 
{\it Statistical Physics II} (Pergamon Press, Oxford).
%
\bibitem{GAP1} G.Y. Chitov and A.J. Millis, preprint.
%
\bibitem{Sher93} P. B{\'e}nard, L. Cheng, and A.-M. S. Tremblay,
Phys. Rev. B {\bf 47}, 15217 (1993).
%
\bibitem{Maeno} Y. Maeno, {\it et al}, 
J. Phys. Soc. Jpn. {\bf 66}, 1405 (1997).
%
\end{references}
\end{document}